\documentclass[conference]{IEEEtran}
\usepackage{cite}
\usepackage{tikz}
\usepackage{amsmath}
\usepackage[numbers]{natbib}
\usepackage[normalem]{ulem}

\usepackage{algorithm}
\usepackage[noend]{algpseudocode}
\usepackage{xspace}
\usepackage{lipsum}
\usepackage{tabularx}
\usepackage{subfig}
\usepackage{graphicx}
\usepackage{url}
\usepackage{titlesec}

\usepackage{flushend}

\usepackage{textcomp}

\newcommand{\algname}{SCL\xspace}

\renewcommand{\paragraph}[1]{\vspace{6pt}\noindent\textbf{#1}}

\titlespacing*{\section}{0pt}{1.5ex}{1ex}
\titlespacing*{\subsection}{0pt}{1ex}{0ex}
\begin{document}

\pagestyle{plain}

\title{\algname: A Secure Concurrency Layer For Paranoid Stateful Lambdas}
\date{}
\author{
\IEEEauthorblockN{
Kaiyuan Chen, 
Alexander Thomas,
Hanming Lu,
William Mullen,
Jeff Ichnowski,
}
\IEEEauthorblockN{
Rahul Arya,
Nivedha Krishnakumar,
Ryan Teoh,
Willis Wang,
Anthony Joseph,
John Kubiatowicz
}
\IEEEauthorblockA{University of California, Berkeley}
\IEEEauthorblockA{\{kych, alexthomas, hanming\_lu, wmullen, jeffi, rahularya, nivedha, ryanteoh, williswang, adj, kubitron\}@berkeley.edu}
}

\maketitle

\thispagestyle{plain}

\begin{abstract}





We propose a federated Function-as-a-Service (FaaS) execution model that
provides secure and stateful execution in both Cloud \emph{and} Edge
environments. The FaaS workers, called \emph{Paranoid Stateful
Lambdas} (PSLs),
collaborate with one another to perform large parallel
computations.  We exploit cryptographically hardened and mobile bundles of data,
called DataCapsules, to provide persistent state for our PSLs, whose
execution is protected 
using hardware-secured TEEs. To make
PSLs easy to program and performant, we build the familiar Key-Value Store
interface on top of DataCapsules in a way that allows amortization of cryptographic
operations.  We demonstrate PSLs functioning in an edge environment running on a group of Intel NUCs with SGXv2.



As described, our \emph{Secure Concurrency Layer} (\algname), provides
eventually-consistent semantics over written values using untrusted and
unordered multicast.  All \algname communication is
encrypted, unforgeable, and private.
For durability, updates are recorded in replicated
DataCapsules, which are append-only
cryptographically-hardened blockchain with confidentiality, integrity,
and provenance guarantees.  Values for inactive keys are stored in a
log-structured merge-tree (LSM) in the same DataCapsule.  \algname
features a variety of communication optimizations, such as an efficient
message passing framework that reduces the latency up to 44x from the
Intel SGX SDK, and an actor-based cryptographic processing architecture
that batches cryptographic operations and increases throughput by 81x.

\end{abstract}


\section{Introduction}
Distributed computing uses workers on multiple hosts to jointly run a
single task. Existing Function-as-a-Service (FaaS) providers, such as
AWS Lambda \cite{lambda}, have pushed distributed computing to an
extreme: users can launch hundreds or thousands of distributed workers
concurrently. Some FaaS implementations, such as
Cloudburst~\cite{sreekanti2020cloudburst}, even support
\textit{stateful} executions, in which distributed workers share storage
with one another.  At this time, the serverless FaaS model has become
quite popular for a wide variety of applications.

\begin{figure}
    \centering
    \includegraphics[width=0.8\linewidth]{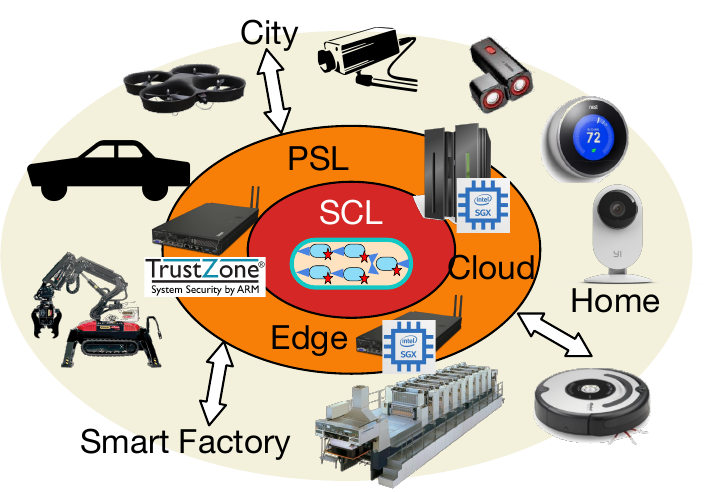}
    \caption{Paranoid Stateful Lambdas (PSLs) exploit pervasive and
      abundant computational resources from the edge and cloud, thus
      enabling a universal plane of secure and stateful in-enclave
      execution. The Secure Concurrency Layer (SCL), provides coherent
      and eventually-consistent semantics over data writes between PSLs,
      while ensuring durability using network-embedded DataCapsules.}
    \label{fig:overview}
\end{figure}

\emph{Edge computing}, in contrast to cloud computing, exploits
resources at the edge of the network, presenting a variety of
opportunities for low-latency, high-bandwidth communication, lower
energy usage, and better privacy.  It is arguably the next major computing
paradigm after cloud computing~\cite{gartner2017edge}.  Providing a
stateful, serverless model of access to resources at the edge seems
ideal for a variety of emerging IoT and robotic applications, since it is
directly compatible with the paradigm of on-the-fly allocation of
compute and storage resources by mobile devices as they transit regions
on the edge of the
network~\cite{ICRAFogRobotics2019,ISRRMotionSegmentation2019,ICRAJeffIcknowski2020}.

Unfortunately, general-purpose Edge computing presents a number of
challenges~\cite{shi2016edge,garcia2015edge} and is thus \emph{not}
widely used by existing FaaS implementations.  One huge challenge is that the edge environment is often not as trustworthy as the cloud;
resources at the edge of the network may be owned and maintained by
novice users or malicious third parties.  Furthermore, \emph{physical}
security is less prevalent in edge environments, leading to a variety of
physical attack vectors.  Compromised devices, while appearing
legitimate, could steal information, covertly monitor communication, or
deny service.  Even worse, malicious entities could seek to alter
information in subtle ways that are not immediately obvious, but which
corrupt edge applications in damaging or even dangerous ways.
Application writers often attempt to ``roll their own'' data protection
in ad-hoc and sometimes buggy ways, leading to data breaches and
security violations.  Clearly, the lack of a standardized approach to
both protect \emph{and} easily utilize information on the edge hinders
exploitation of edge computing resources.


\paragraph{Paranoid Stateful Lambdas:}
In this paper, we introduce the
first FaaS execution service that enables secure \emph{and} stateful execution
for both cloud and edge environments, while at the same time being easy
to use.  The FaaS workers, called \emph{Paranoid Stateful Lambdas}
(PSLs), can collaborate to perform large parallel
computations that span the globe and \emph{securely} exploit resources
from many domains.  See Figure~\ref{fig:overview}.  We provide an
easy-to-use data model and automatically manage cryptographic keys
for compute and storage resources.

We take a two-fold approach to supporting PSLs. First, we
exploit trusted execution environments (TEEs), such as those provided by Intel
SGX~\cite{arnautov2016scone,costan2016intel} and ARM TrustZone.  TEEs
provide confidentiality and integrity of executed functions while also
providing strong isolation of data, computation, and cryptographic
assets from the untrusted kernel or hypervisor.  Through attestation, a
multi-PSL application can be served by an untrusted, third-party service
provider.

\begin{figure}
    \centering
    \includegraphics[width=0.95\linewidth]{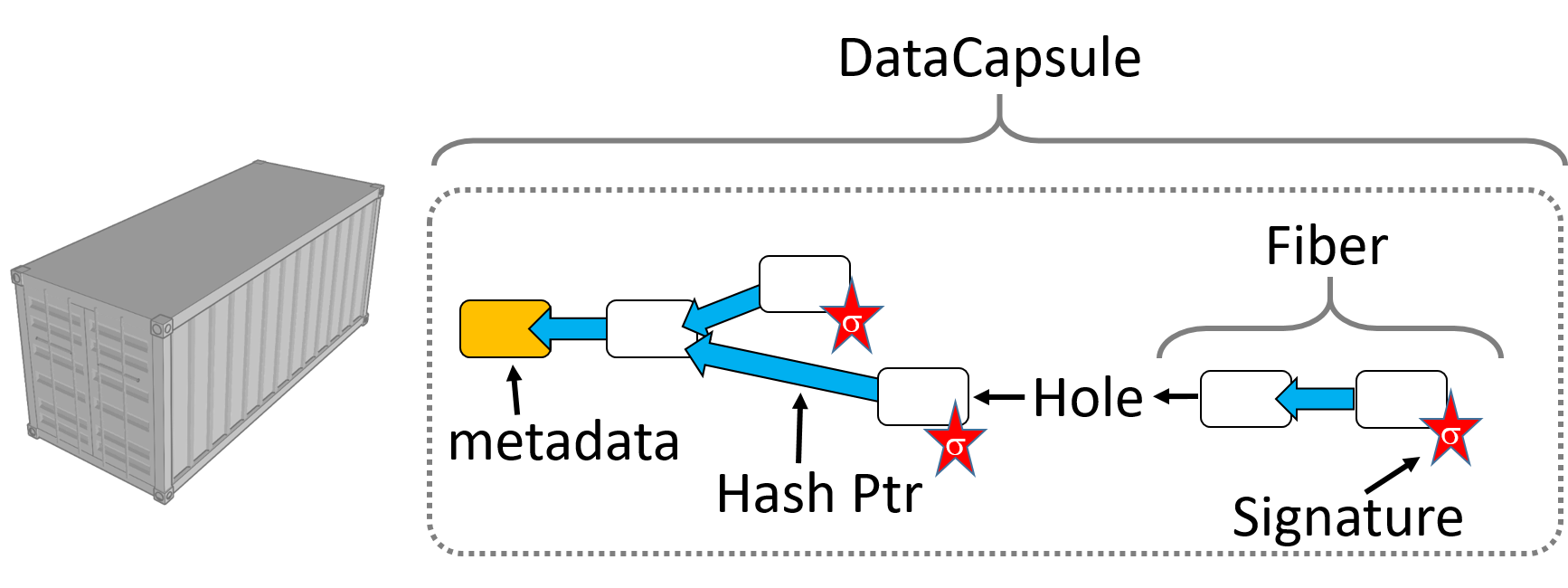}
    \vspace*{-0.05in}
    \caption{DataCapsule: a \emph{cohesive} representation of
      information analogous to a shipping container.  Internally, it is
      an ordered collection of immutable \emph{records} that are linked
      to each other and signed.  Its unique, \emph{routable} identity is
      derived from hashes over metadata including ownership credentials.}
    \label{fig:datacapsulecontainer}
\end{figure}

Second, we package persistent state in cryptographically hardened
bundles of data called \emph{DataCapsules}~\cite{mor2019global}. Each
DataCapsule is a ``blockchain~\cite{zheng.z2017} in a box,''
exploiting a standardized metadata format to guarantee provenance,
integrity, and privacy via cryptography.  With 
append-only semantics, DataCapsules provide a permanent audit-trail of
operations on their contents, thus allowing undo-like operations,
multi-version support, and mediation in the presence of malicious
failure.  See Figure~\ref{fig:datacapsulecontainer}.  While DataCapsules
can be embedded in any network storage environment, they are
particularly powerful when combined with a data-centric network such as the
\emph{Global Data Plane} (GDP)~\cite{mor2019global}, which allows
DataCapsules to be stored, migrated, and interacted with anywhere in the
network.  In this paper, we treat the DataCapsule service as a black box
provided by the underlying infrastructure.

 
This combination of TEEs (for active computation and data in use) and
DataCapsules (for data at rest or in motion) provides a powerful
combination that enables secure, stateful computation in insecure
environments.  While DataCapsules by themselves provide a standardized
way to encapsulate and protect information as it moves within the
network (leading to a global, federated data service), we ease
the burden of PSL programmers by presenting them with a familiar
key-value store (KVS) interface.  This is implemented on top of DataCapsules via a
protected ``runtime system'' we call a Common Access API, or (CAAPI).
Consequently, communicating PSLs may interact via shared keys in the
key-value store.



\paragraph{Performance Challenges:}While security is one of
our primary motivations for implementing the PSL framework, we
also wish to enable \emph{high performance} parallel computation
using PSLs.  This goal is hindered in multiple ways: First, the
distributed nature of PSL-based parallelism leads to a need for
\emph{relaxed consistency} for most writes in our KVS.  We
discuss how to implement an eventually-consistent model for
\texttt{put()} operations that allows interacting PSLs to operate with
independence from one another while still detecting denial of service
attacks and bounding the maximum write propagation delay over an
unordered and untrusted network. Our mechanism further enables a
release-consistent locking scheme~\cite{gharachorloo1990}.

Second, all communication between collaborating enclaves must be
encrypted and signed to prevent malicious parties from forging,
corrupting, or observing such communication. This \emph{security tax}
can be significant if not mitigated through batching and suppression of
locally overwritten updates.  We show how our relaxed consistency
implementation permits a variety of cryptographic optimizations.

Third, the strong isolation provided by TEEs is a double-edged sword:
while shielding in-enclave applications from external malicious parties,
it imposes a strong impediment to communication across the enclave
barrier.  The common communication approach \cite{costan2016intel}
involves hardware-specific attestation and complicated key exchange
protocols for one-to-one communication, the complexity only increasing
with larger enclave group sizes.  Even crossing the enclave
barrier on a local node using the popular SGX container framework
(GrapheneSGX~\cite{tsai2017graphene}) can exhibit horrendous overhead,
combining an expensive context switch with byte-wise data
copying\footnote{The standard \emph{system call}
facility for SGX incurs between 8,000 and 20,000 cycles for an
\emph{ecall} and takes 8,000 cycles for an \emph{ocall}.}.
Our approach to speeding up communication exploits the standardized
DataCapsule format (which protects information) combined with heavy
optimization of communication across the enclave barrier and a fast but
untrusted multicast tree for communication.  

\paragraph{The Secure Concurrency Layer:}
Much of our communication innovations are embodied in the eponymous
\emph{Secure Concurrecy Layer} (\algname), one of the primary topics of this
paper. \algname is an in-enclave cache manager that securely and
efficiently relays data between multiple enclaves while providing
well-formed update semantics.
In our system, a given PSL interacts with remote PSLs by issuing KVS
\texttt{put()} operations to its own local cache.  \algname translates
these write operations into encrypted and signed update records
compatible with the underlying DataCapsule.  The updates are then
propagated to other enclaves as well as the network-embedded DataCapsule
(for durability) over an untrusted and unordered network multicast tree.
\algname provides eventual consistency semantics over the written
values, but enforces epoch-based resynchronization for
liveness\footnote{Our system utilizes a Log-Structured Merge(LSM) tree
to efficiently store idle Key-Value pairs, namely those not currently in
PSL caches.}.  \algname also features various performance
optimizations. For example, \algname uses a circular buffer based
message passing design, which passes messages across secure enclave
boundary 44x faster than using standard \texttt{send} ecalls. To
parallelize the cryptographic computations, such as encryption, hashing,
and signing, \algname uses an actor-based architecture for computing the
DataCapsule's headers. When combined with batching, these optimizations
increase throughput by 81x over the unoptimized baseline.


We design and implement the PSL FaaS infrastructure using
\algname. PSL-enabled worker nodes can run
directly on top of \algname by static linking or dynamic script
interpretation\footnote{In the future, we hope to support dynamic
linking of PSL binaries residing in DataCapsules.}.
To bootstrap secure enclaves with appropriate cryptographic identities,
we design a key management scheme inspired by the Bitcoin
wallet~\cite{Antonopoulos2017} and an optimized attestation protocol.
Unlike previous works \cite{weisse2017regaining,kim2019shieldstore} that
only support Intel SGX and assume SGXv1,
we implement \algname on Asylo~\cite{asylo,asylosocket}, a hardware-agnostic framework that
allows \algname to run on most mainstream TEE hardware.
The result is a third-party service running on
the edge that can satisfy on-the-fly requests to securely execute PSL
applications using compute and storage resources embedded in the edge
environment.

We claim the following contributions in this paper: 
\begin{itemize}
    \item \textbf{Paranoid Stateful Lambdas (PSLs):} We introduce the notion of
      Paranoid Stateful Lambdas and show the design and implementation of
      our PSL execution environment. 
    \item \textbf{Separation of State and Computation:} We
      propose to use DataCapsules as the ground-truth vehicle for
      communication among different types of secure enclave hardware with
      confidentiality, integrity, and provenance guarantees.
    \item \textbf{\algname KVS:} We design, implement and evaluate
      \algname, a secure and eventually-consistent replicated KVS that
      facilitates inter-enclave communication and bounds maximum write
      latency while mitigating denial of service.  We implement
      associated key distribution and attestation protocols.
    \item \textbf{Communication Optimizations:} We reduce and amortize
      the communication and cryptographic overhead by rearchitecting the
      cryptographic pipeline and designing a circular buffer based
      message passing mechanism.
\end{itemize}



\section{Background}
\subsection{Secure Enclaves}
\label{sec:sgx}
Our design does not assume specific secure enclave hardware or a set of
supported instructions; we only require the trusted hardware to have
semantics for memory protection and attestation.  Here, we introduce
Intel Software Guard Extensions (SGX)~\cite{costan2016intel} due to its
wide adoption.  SGX allows users to create a secure, isolated
environment protected from the privileged host OS, hypervisor, or any
hardware devices connected to the host. SGX protects against physical
adversaries and uses a hardware Memory Encryption Engine (MEE) to
guarantee the confidentiality and integrity of enclave memory. All
enclave memory must occupy a specific section of memory in the enclave
page cache (EPC). If an EPC page is evicted, it is encrypted and stored
onto the disk. An EPC page that is loaded back into memory is integrity
checked and decrypted.

The host OS is still responsible for mapping page tables and allocating
memory, but subsequent memory accesses are checked by SGX. SGX ensures that enclave memory can only be accessed by the specific enclave the page is allocated to when walking the page table. In SGXv1, the EPC is a limited resource and has a static limit of 128MB shared across all enclaves, while SGXv2 dynamically allocates an EPC that can be oversubscribed by multiple secure enclaves.

SGX can also verify the identity of an application running inside a
secure enclave. Intel allows for an attestation report of an enclave
to be generated. This
report includes a measurement of the code and data sections of the
application binary signed by a hardware root of trust
and can be verified through \textit{Intel's Attestation Services} (IAS).



\begin{figure*}
    \centering
    \includegraphics[width=0.7\textwidth]{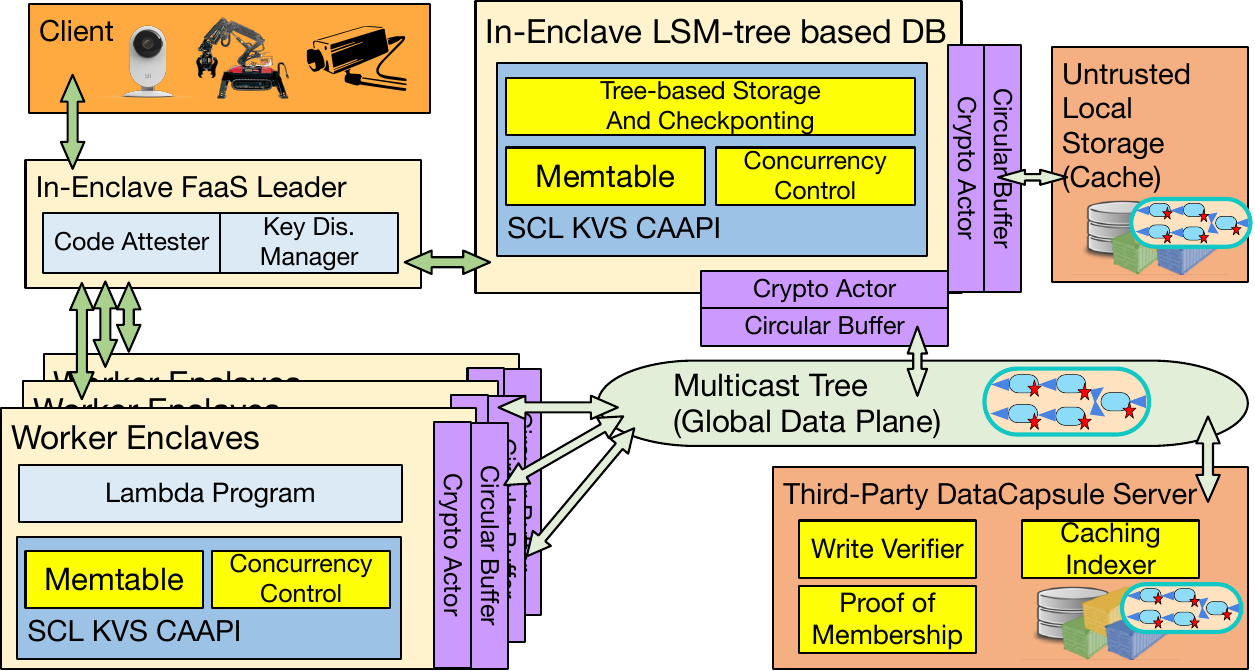}
    \vspace*{-0.05in}
    \caption{An Overview of Paranoid Stateful Lambdas. The client first
      attests the integrity of the Secure FaaS Leader code. The FaaS
      Leader distributes cryptographic keys and dispatches FaaS task
      (Lambda Program) to other attested Worker Enclaves.  Workers
      perform stateful FaaS computation by communicating with other
      Workers on GDP-based multicast tree. SCL enables secure
      communication by providing a replicated KVS abstraction. For
      persistence, PSL uses an in-enclave LSM-tree based database that
      stores inactive keys. In this paper, we focus on the security and
      communication aspect of PSL.}
    \label{fig:full_system_diag}
\end{figure*}

\subsection{DataCapsules and the Global Data Plane}
\label{subsec:datacapsulesGDP}
We briefly discuss the benefits of adopting
DataCapsules~\cite{mor2019global} as the underlying
storage objects for PSLs.  DataCapsules are bundles
of data containing data chunks (``records''), along with cryptographic
relationships between these records (\emph{i.e.} hashes) and proofs of
membership and/or provenance for these records (\emph{i.e.} signatures);
see Figure~\ref{fig:datacapsulecontainer}.  DataCapsules have an
\emph{owner}, which is a public/private key pair.  Anyone with the
private owner key can add records to a DataCapsule but cannot modify
existing records.  Consequently, write operations are append-only
and must be accompanied by a
\emph{signature} from the DataCapsule owner.  Read operations return a
\emph{proof of membership}, consisting of a signature and a chain of
hashes along with data.  Thus, it is possible to perform secure
operations on remote DataCapsules embedded within the network.

The Merkle-tree structure of hashes and signatures within a DataCapsule
provides two major benefits: \emph{First}, it is a Conflict-Free
Replicated Data Type (CRDT).  This means any two partial
DataCapsule replicas can be easily synchronized by simply taking the
union of records between them; the resulting tree is uniquely defined by
the backlinks (hashes).  \emph{Second}, it prevents malicious parties
from forging records or corrupting existing records; the worst that a
malicious party could do with a DataCapsule is executing a
\emph{freshness attack} by denying the presence of recent records.
Freshness attacks can be prevented or mitigated in a variety of ways,
including replication, periodic timestamping of records, and caching of
pointers to the most recent records.  For our PSL infrastructure, we
start by requesting the most recent records from a trusted \emph{service
provider}, then maintain the most recent ``wavefront'' of signed records within
our active enclaves.  As a result of their hardened nature, DataCapsules
can migrate to the edge and benefit from its storage and networking resources. 

Since DataCapsules can be viewed as secure logs, they can encapsulate a
wide variety of storage ``patterns,'' such as key-value stores (in this
paper), filesystems, data streams, and databases.  All that is required
to implement such patterns is a layer of software, called a \emph{common
access API} (CAAPI), that accepts standard user requests (\emph{e.g.} POSIX
filesystem requests) and translates them to operations on the
underlying DataCapsule.  Such CAAPIs run in secure enclaves,
since they need access to cryptographic keys to produce signatures and
to encrypt/decrypt information over the DataCapsule API.

In this paper, we assume that DataCapsules reside in some network server
that is able satisfy DataCapsule read and write operations.  However,
the true power of DataCapsules is revealed in the context of a
data-centric network such as the Global Data Plane
(GDP)~\cite{mor2019global}.  Each DataCapsule has a unique 256-bit
identity derived from a hash over the public owner key and other
metadata.  The GDP can route messages to a DataCapsule \emph{using its
identity} rather than a location (\emph{i.e.} an IP address).  Thus, a
data client can send reads and writes to a DataCapsule without knowing
its location\footnote{When multiple DataCapsules exist with the same
identity, they are assumed to be equivalent; thus the GDP will try to
route queries to the ``closest'' equivalent DataCapsule.
Replication thus provides a mechanism for content distribution,
providing a cryptographically hardened form of CDN with well-defined,
in-network update semantics, unlike alternatives such as
NDN~\cite{zhang2014named}.}.  Thus, with the GDP, PSLs could launch anywhere
and access their data simply by possessing (1) the unique identity of
the DataCapsule containing its data, (2) the cryptographic ownership and
encryption keys for the DataCapsule, and (3) a connection into the GDP.

\section{Paranoid Stateful Lambda}
\label{sec:psl}
Paranoid Stateful Lambdas (PSLs) provide unifed access to the
computation and storage resources of the cloud and edge. They provide
access to the abundance of edge servers which have better locality and
lower latency than would be available with cloud-only environments.  The
serverless abstraction enables applications to be transparent about the
underlying infrastructure.

\paragraph{Paranoid:} PSL allows clients to launch a scalable number of distributed workers (\emph{i.e.} \textit{Lambdas}) on both cloud clusters and edge servers. Recognizing that servers on the cloud and edge may come from mutually distrustful service providers,  PSL executes all the privacy-sensitive programs in secure enclaves, guaranteeing the confidentiality and integrity of all executions. 

For the threat model, PSL adopts the typical "cloud/edge attackers" who can listen and tamper with any communications or computations. For example, the attack may come from a compromised operating system kernel or a malicious staff member, both situations in which the attacker has full control over the system. \algname guarantees the confidentiality, integrity, and provenance of any data in execution and in transit. 
The trusted computation base (TCB) of SCL is limited to the processor chip, PSL code, and sandboxed application code running in an enclave, which explicitly excludes the operating system managed by the cloud provider. The design of SCL guards against message replay attacks and detects DDoS attacks at a granularity of a user-defined time interval (epoch). However, PSL does not guarantee against side-channel attacks, given that  Intel SGX suffers from various side-channel vulnerabilities~\cite{10.1145/3052973.3053007, sgxsdk, tsgx}. However, there are various techniques~\cite{10.1145/3052973.3053007,216033, tsgx,10.1145/2897845.2897885} proposed to mitigate the risk of side channel attacks.

\paragraph{Stateful:} Beyond other secure FaaS implementations \cite{alder2019s}, PSL supports \textit{stateful} execution of distributed workers, meaning that one in-enclave worker is able to communicate with workers in other enclaves or even workers that will be executed in the future \cite{sreekanti2020cloudburst}. Statefulness has already become a necessity in many popular FaaS applications: for example, ExCamera~\cite{fouladi2017encoding}, numpywren~\cite{shankar2018numpywren}, mplambda~\cite{ICRAJeffIcknowski2020}. 

In order for Lambdas to be Paranoid and Stateful, PSL consists the following main components: 
(1) \textbf{Secure Concurrency Layer (SCL)}: enables secure communication between multiple enclaves, (2) \textbf{In-Enclave LSM-tree based DB}: provides persistence and durability of the DataCapsule, (3) \textbf{PSL Secure FaaS}: securely attests SCL, distributes cryptographic keys, and dispatches tasks to Worker Enclaves, and (4) \textbf{Global Data Plane} \cite{mor2019global}: provides global routing infrastructure. 

\paragraph{Secure Consistency Layer:} In designing PSL, we recognize the
need to have a \textbf{secure} layer that allows enclaves to communicate
and \textbf{concurrently} share objects.  
This layer provides security and consistency semantics for transient
messages over untrusted and unordered multicast.
Consequently, distributed worker programs can use this layer as a form
of shared memory, and PSL as a whole can use this layer to dispatch program scripts
and coordinate idle secure enclaves. An analogy to this layer is
BigTable for Google or Dynamo for Amazon, infrastructure which provides
a KVS layer as foundational communication abstraction to higher level
applications.

To enhance performance, we designed an eventually-consistent replicated KVS
that presents a shared memory view to all the secure enclaves connected
to the same network multicast tree.
If an enclave makes KVS updates to the local cache, the
changes will be propagated to all other secure enclaves by
broadcast. The secure enclaves maintain the same copy of memory cache.
\algname partitions the KVS into a memtable that fits in main memory,
and PSL has a Log-Structured Merge (LSM) tree inspired by RocksDB
\cite{rocksdb} that stores inactive keys.

\section{SCL Design}

\subsection{Overall Architecture}

\begin{figure}
    \centering
    \includegraphics[width=0.85\linewidth]{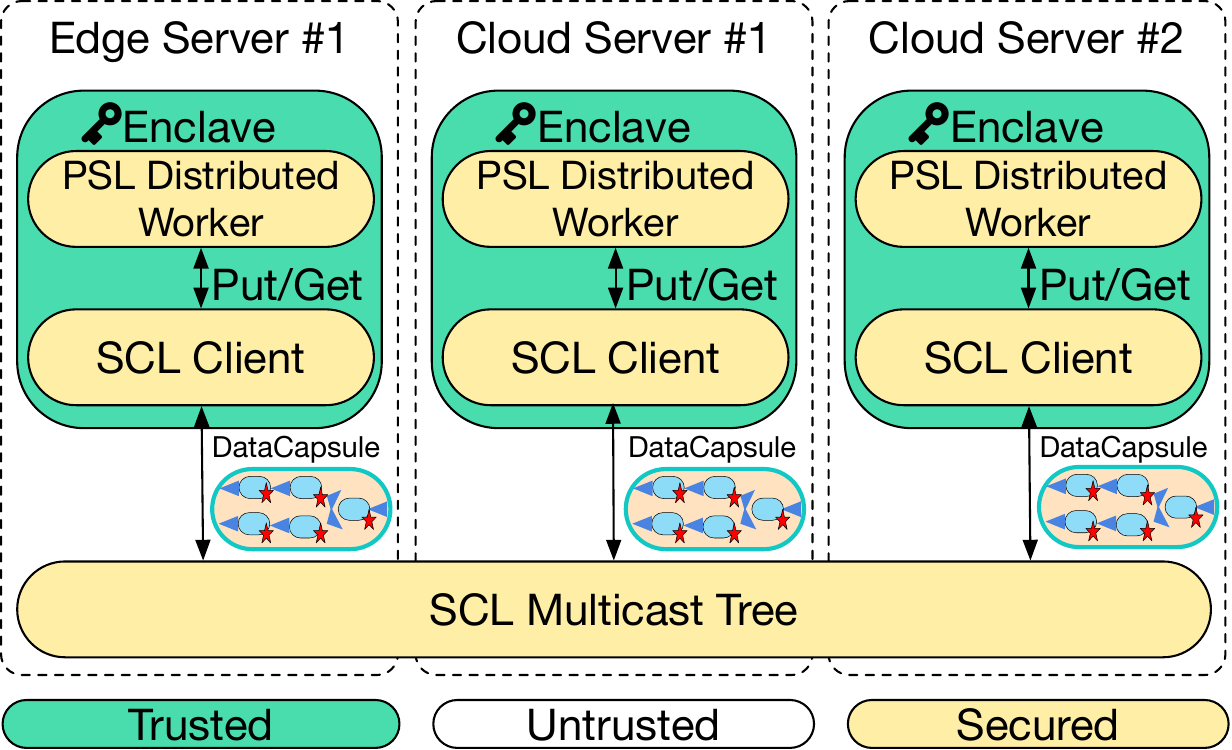}
    \caption{The architecture of \algname.  In-enclave workers communicate with each other by interacting with PSL using \texttt{put()} and \texttt{get()} operations. The KVS updates are propagated by \algname with a secure data structure called a DataCapsule.}
    \label{fig:architecture}
\end{figure}

The essence of \algname is that every secure enclave replicates a
portion of the underlying DataCapsule (Section \ref{sec:dc}), namely the
portion dealing with active keys.  This portion can be thought of as a
write-ahead log. By allowing this log to branch, we free workers to be
independent of one another for periods of time.  The intuition is that
we arrange these temporarily divergent histories to include sufficient
information to provide well-defined, coherent, and eventually-consistent
semantics. We do so while allowing updates to be propagated over an
insecure and unordered multicast tree.



Figure \ref{fig:architecture} provides an overview of \algname's
architecture. Each enclave maintains an in-memory replicated cache
called a \emph{memtable}. All \texttt{put()} operations are placed into
the local memtable, timestamped, and linked with previous updates before
being encrypted, signed, and forwarded via multicast.  The
code which performs these operations is a CAAPI that
provides the KVS interface on top of the DataCapsule storage.

The DataCaspule record appends are propagated by a network
\emph{multicast tree}. If an enclave receives a record to append, it
verifies the cryptographic signatures and hashes, and merges the record
with its own replica of the DataCapsule. The merge does not require
explicit coordination due to the CRDT property of the DataCapsule hash
chain, but the hash chain is periodically synchronized to bound the
consistency. DataCapsule changes are also reflected to the memtable with
the eventual consistency semantics. \algname enables fast KVS read
queries because of its shared memory abstraction, so all \texttt{get()}
operations can directly read from the enclave's local memtable without
querying other nodes.

\subsection{DataCapsule Contents For SCL}
\label{sec:dc}

\begin{figure}
    \centering
    \includegraphics[width=0.7\linewidth]{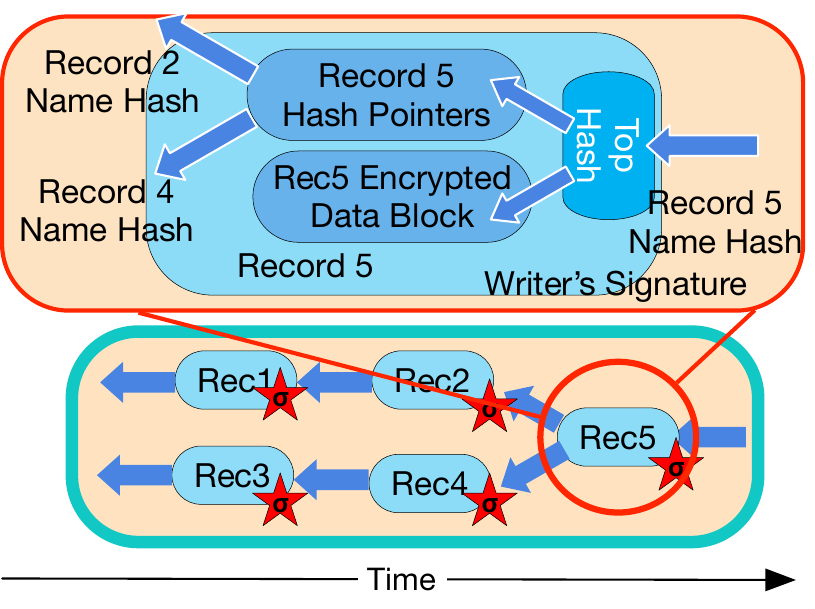}
    \vspace*{-0.1in}
    \caption{Every DataCapsule record is encrypted, signed, and appended to the DataCaspule by including one or more hashes to previous records.}
    \label{fig:dc_struct}
\end{figure}



Figure \ref{fig:dc_struct} shows a visual representation of DataCapsule
contents for SCL. Each record contains one or more SHA256 hash pointers to previous records, encrypted data, and a signature.
Multiple previous hash pointers occur during \emph{epoch-based
resynchronization}, which we discuss shortly. The data block of a record
includes \texttt{SenderID}, a unique identifier of the writer,
\texttt{Timestamp}, that indicates the sending time of the record, and
\texttt{Data}, the actual data payload. In SCL, \texttt{Data} is an
AES-encrypted string that contains the updated key-value pairs.  The
record contains a \texttt{Signature} signed on the entire record with
Elliptic Curve Digital Signature Algorithm (ECDSA) using the private key
of the writer.

\begin{figure}
    \centering
    \includegraphics[width=0.92\linewidth]{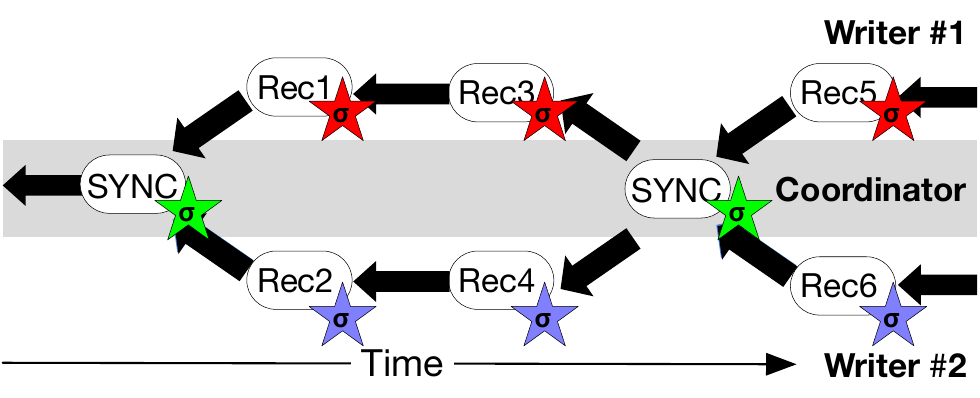}
    \vspace*{-0.1in}
    \caption{DataCaspule hash chain with two concurrent writers (\#1 and \#2) using epoch-based resynchronization. The writers write to the same hash chain and the coordinator uses a SYNC report to aggregate the DataCapsule updates from each writer. }
    \label{fig:sync}
\end{figure}

\subsection{Memtable}
\label{sec:kvs}
The Memtable KVS is an in-enclave cache of the most recently updated
key-value pairs.  The key-value pairs are stored in plaintext, as the confidentiality and
integrity of the memtable are protected by the secure enclave's EPC.
In-enclave distributed applications communicate with other enclaves by
interacting with the memtable using the standard KVS interface:
\texttt{put(key,value)} to store a key-value pair and \texttt{get(key)}
to retrieve the stored value for a given key.

\algname communicates with other enclaves to replicate the memtable
consistently across enclaves.
Local changes to the memtable are propagated to other enclaves and
updates received from other enclaves are reflected in the local
memtable.
When receiving an update from
other enclaves, the memtable performs verification and decryption
before putting the key-value pair into the memtable. To achieve
coherence, updates received from remote enclaves are only placed in the
local cache if they have later timestamps (see below).



\subsection{Consistency} 


\algname guarantees \emph{eventual consistency} of values associated
with each key, namely that if there were no further updates to a
specific key $k$, then \texttt{get($k$)} from each of the in-enclave
memtables should return the latest value.  In addition, \algname
guarantees the property of \emph{coherence} among values, namely that no
two enclaves will ever observe two updates to a given key in different
orders. Since updates are propagated over an unorder multicast tree, 
\algname needs to order the DataCapsule updates in the memtable, and
reject updates that are causally earlier than the ones already in
the memtable.  Naive approaches may lead to undesirable outcomes: for
example, replacing the memtable's value whenever a new record arrives
leaves the memtable in inconsistent state.

To decide the order of the updates, \algname uses a Lamport logical
clock to associate every key-value pair with a logical
timestamp. Without relying on the actual clock \textit{time}, \algname
increments a local Sequence Number (SN) whenever there is an update to
the local memtable. When receiving a new DataCapsule record, it also
synchronizes the SN with the received SN in the record by $SN \gets
max(local SN, received SN)+1$.

Although a vector clock is usually deemed as an upgrade to Lamport
logical clocks by including a vector of all collected timestamps,
\algname uses a Lamport clock because a DataCapsule already carries
equivalent versioning and causality information, and a vector clock
introduces additional complexity and messaging overhead to the
system. The reason for not using the actual timestamp from the operating
system is that getting such timestamp costs more than 8,000 CPU cycles
due to enclave security design \cite{weisse2017regaining}. Getting
trusted hardware counters from RDTSC and RDTSCP instructions is also
expensive ($60$-$250$ ms)~\cite{bailleu2019speicher} only supported by SGX2.

\subsection{Epoch-based Resynchronization}
\label{sec:consistency}

For performance,
\algname branches the DataCapsule data
structure by letting every enclave write to its own hash chain. Every
append includes a hash pointer to the previous write from the same
writer, instead of the previous write across all enclaves. To merge the
hash chains maintained by each writer, \algname uses a synchronization
(SYNC) report as a rendezvous point for all DataCapsule branches. The resultant DataCapsule hash chain is structured in a diamond
shape like Figure \ref{fig:sync}.  All writers' first writes include a
hash pointer to the previous SYNC report, and the next SYNC report
includes a hash pointer to the last message of every writer. The SYNC
report is useful in the following ways:
\begin{itemize}
\item\textbf{Detecting the freshness of a message:} Every message
      includes a monotonically increasing SYNC report sequence number,
      which is incremented when a new SYNC report is generated. As a
\item\textbf{Fast Inconsistency Recovery:} After an enclave
      receives a SYNC report, it can use the hash pointers to backtrack
      to previous messages from the same writer until it reaches the
      last SYNC report. It detects a message is lost if a hash pointer
      cannot be recognized during the backtrack.
\end{itemize}

The usage of SYNC reports establishes a notion of epoch-based
resynchronization, a tradeoff between synchronization overhead and
consistency. With epoch-based resynchronization, the user defines a
synchronization time interval called an epoch. Between the epochs, the
enclaves use timestamps to achieve coherency and eventual consistency.
At the end of an epoch, enclaves cross-validate
their own DataCapsule replica with the SYNC report generated by a
special Coordinator Enclave (CE).

\subsection{Multicast Tree}
\label{sec:mcast}

We conclude this section by showing the overall structure of the
multicast tree. Multicast enables one node on the multicast tree to
communicate with multiple nodes by routers. Routers receive the message
and re-broadcast the message to multiple nodes or routers. The overall
structure forms a multicast tree. The tree-like structure extends the
scalability that allows more than one router to handle the
communication. \algname is agnostic to the multicast tree topology, as
long as a node on the tree publishes the message, and all the rest of
the nodes can receive that message.  As a result, we can abstract
\algname's multicast tree structure as a plane where worker enclaves publish DataCapsule updates
through multicast routers, the third party durability storage can log
all the messages and third-party authenticators can verify the validity
of the DataCapsule hashchain.


\subsection{Duraibilty and Fault Tolerance}
\label{sec:fault}
We discuss the durability semantics of \algname, and how we use it to store inactive keys.  We also discuss \algname when facing multiple types of
failures. 




\paragraph{SCL Durability with DataCapsule:}
Any secure enclave in \algname may fail by crashing or losing its network
connection, causing it to fall behind or even leave its in-memory states
inconsistent. SCL is durable if a such enclave can recover all of the
in-memory states (\emph{i.e.} the memtable) consistently and catch-up with the
on-going communication. Due to the equivalence of the DataCapsule hash chain
and the memtable shown in Section \ref{sec:kvs}, the durability of SCL
is achieved by making the DataCaspule hash chain persistent. An analogy to
SCL's durability is Write-ahead Logging(WAL) in many databases,
which logs an update persistently before committing to the permanent
database. In SCL, DataCapsule is the append-only log that records the
entire history of the memtable, which a crashed enclave can use to
recover. The crashed enclave can merge its local inconsistent
DataCaspule hash chain with the persistent DataCapsule received from
other components, either an in-enclave LSM tree DB or DataCapsule servers. The CRDT property of DataCapsule guarantees the consistency
of the hash chain after the merge.

\paragraph{DataCapsule Replication: } A DataCapsule replica may fail by crashing or network partitioning, resulting in service interruption by SCL. A DataCapsule replica may also be corrupted and lose partial or the entire SCL data permanently. To ensure durability and availability of DataCapsule replicas, we implement a continuous DataCapsule replication system that uses write quorum to tolerate user-defined \textit{f} replica failures or network partitioning in the system without disturbing the PSL computation.

\paragraph{Coordinator Failure:}  The failure of the coordinator only
influences the resynchronization interval, but does not influence the
strong eventual consistency given by the CRDT property of DataCapsules
and the logical timestamp of the memtable. However, one can use multiple
read-only shadow coordinators to improve the fault tolerance and to
remove the single point of failure. If multiple consistency coordinators are
on the same multicast tree, one consistency coordinator can actively
send RTS broadcasts to the multicast tree. Other consistency
coordinators remain in shadow mode until a SYNC report is not sent for
an extended period of time.

\subsection{CapsuleDB}
CapsuleDB is a key-value store inspired by LSM trees and backed by DataCapsules.  It is built to specifically take advantage of the properties of DataCapsules to provide long-term storage of large amounts of data as well as accelerate PSL recovery beyond reading every single record in the DataCapsule.  Figure~\ref{fig:full_system_diag} shows how CapsuleDB fits into the PSL framework with its separate enclave running on behalf of the Worker Enclaves.

CapsuleDB has two main data structures, CapsuleBlocks and indices, as well as its own memtable.  CapsuleBlocks are groups of keys, each of which represents a single record in the DataCapsule backing the database.  It's data storage structure is inspired by level databases such as RocksDB \cite{rocksdb} and SplinterDB \cite{conway2020splinterdb}.  Data is split into \textit{levels}, each with increasing size.  In CapsuleDB, Level 0 (L0) is the smallest while subsequent levels L1, L2, and so on increase by a factor of ten each time.  Each level is made up of CapsuleBlocks.  In L0, each block represents a memtable that has been filled and marked immutable.  Blocks in lower levels each contain a sorted run of keys, such that the keys in the level are monotonically increasing.  

The index manages which CapsuleBlocks are in each level, the hashes of each block, and which blocks contain active data.  The index also acts as a checkpoint system for CapsuleDB, as it too is stored in the DataCapsule.  While the main purpose of storing the index in the DataCapsule is to ensure CapsuleDB can quickly restore service after a failure, it has the added benefit that old copies of the indices serve as snapshots of the database over the lifetime of its operation.  


\paragraph{Writing to CapsuleDB} 
CapsuleDB participates in SCL just like the worker enclaves. Consequently, every write is stored into CapsuleDB's local memtable. In this way, it has visibility to the most recent values written by the workers.  Once the memtable fills, it is marked as immutable and appended to the DataCapsule.  The resulting record's hash is then stored in L0 in the index.  If this write causes L0 to become full, the compaction process, described below is triggered, pushing compacted blocks out to the DataCapsule.  

\paragraph{Reading from CapsuleDB} 
Retrieving a value from CapsuleDB is triggered by a {\tt get} operation from one of the worker enclaves when it attempts to find a value that is not stored in its local memtable.  This {\tt get} is routed to CapsuleDB where it begins at the CapsuleDB's memtable.  If the requested key is not found, the request moves to the index for CapsuleBlock retrieval.  Note that once CapsuleDB finds a value for the requested key, it multicasts the result as a {\tt put} on the multicast tree with a timestamp from the blocks---exactly as if it were a worker enclave.

When searching for the most recent value associated with a key, CapsuleDB checks the index associated with L0.  If the key is found after scanning through each block, then the corresponding tuple is returned.  If the search fails, the process is repeated at L1.  However, L1 is sorted, so our search can be performed substantially faster.  In addition, L1 is likely too large to bring fully into memory, especially given the tight memory constraints imposed by some secure enclaves.  As such, only the requested block is retrieved from the DataCapsule.  Again, it is checked to see if the requested KV pair is present.  If not, the same procedure is run at lower levels until it is found, or CapsuleDB determines it does not have the requested KV pair.   
 
\paragraph{Indexing the Blocks} 
The index is at the core of CapsuleDB and serves several critical roles.  Primarily, it tracks the hashes of active CapsuleBlocks to quickly lookup keys.  Whenever a block is added, removed, or modified, the hash mapping is updated in the index.  When compaction, the process of moving old data to lower levels, occurs, the index updates which levels the moved CapsuleBlocks are now associated with.  In this way, CapsuleDB can always quickly find the most recently updated CapsuleBlock that may have a requested key. Further, CapsuleDB keeps a complete record of the history of updates to the KV store, effectively acting similiar to a {\tt git} repository.



Since the index is written out whenever it is modified, the CapsuleDB instance can be instantly restored simply by loading the most recent index.  Then, only the records since the last update need to be played forward to restore the most recent KV pairs that were in CapsuleDB's memtable.  

\paragraph{Compaction} 
Compaction is critical to managing the data in any level-based system.  CapsuleDB's compaction process uses key insights from the flush-then-compact strategy of SplinterDB \cite{conway2020splinterdb} to limit write amplification.  Each level has a maximum size; we say a level is full once the summed sizes of the CapsuleBlocks in that level meets or exceeds the level's maximum size.  This triggers a compaction.  

All writes to CapsuleDB are first stored in the in-memory memtable.  Once the memtable fills, it is marked as immutable and written to L0 as a CapsuleBlock, also simultaneously appending the block as a record to the DataCapsule.  Once L0 is full, compaction begins by sorting the keys in L0.  They are then inserted into the correct locations in L1 such that after all the keys are inserted L1, the level is still a monotonically increasing run of keys.  Any keys in L0 that are already present in L1 would replace that data in L1, since the L0 value and timestamp would be fresher.  Finally, the CapsuleBlocks and their corresponding hashes are written out to the DataCapsule and updated in the index, marking the end of compaction.  



\section{Optimizations}
\label{sec:optimization}
In this section, we discuss optimizations that significantly improve the throughput of \algname.  Because of the high overhead of various cryptographic operations when constructing DataCapsule fields, we propose an actor-based architecture to pipeline the cryptographic operations. The proposed pipeline also enables high-throughput batching and message prioritization. We discuss our circular buffer, a design that efficiently passes messages across application-enclave boundary.

\subsection{Actors and Batching}
\label{sec:actor}
\begin{figure}
    \centering
    \includegraphics[width=0.9\linewidth]{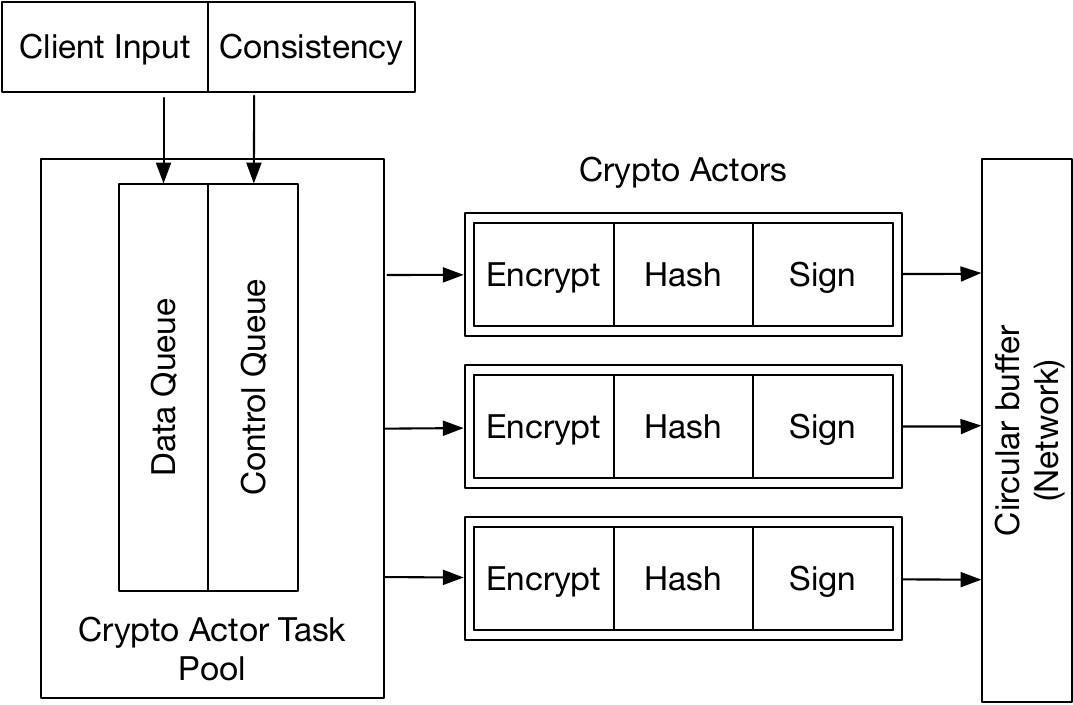}
    \caption{An \algname endpoint with three actors. Both Clients and DataCapsule Consistency coordinator push data and control messages to Crypto Actor task pool. The crypto actors take messages from the queue and perform encryption, hashing and signing. After processing, the crypto actors put the processed message into the circular buffer.}
    \label{fig:multithreading}
\end{figure}

For security, one DataCapsule transaction involves encryption, signing and hashing. These cryptographic operations combined introduce large computational overhead to the critical path when the client issues a \texttt{put}. Because frameworks such as Intel SGX SDK~\cite{sgxsdk} and Asylo~\cite{asylo} do not support async operations, \algname introduces actors to amortize the  overhead.

When the client issues \texttt{put(k, v)}, \algname piggybacks a timestamp $t$ to the key-value pair and pushes the $(k, v, t)$ tuple to a thread-safe Data Queue. The control and coordination messages, such as RTS and EOE, are sent to a thread-safe Control Queue. We call the Data Queue and Control Queue together the Crypto Actor Task Pool. \algname starts multiple threads as crypto actors. These actors take the messages from the task pool, first drawing from the higher priority Control Queue due to latency constraints, and process them into encrypted, hashed, and signed DataCapsule transactions. The generated DataCapsule records are put into the circular buffer and propagated to other enclaves.

\paragraph{Batching:} To optimize the overall throughput and amortize the cost of transmission, \algname can batch multiple key-value pairs in the same DataCapsule record. A crypto actor takes $(k, v, t)$ tuples from the Task Pool. The batch size is the max of the user-preset batch size and the remaining tuples in the task pool. 
The actor serializes all the control messages and key-value tuples into a Comma-Separated Values (CSV) string and feed into cryptographic pipeline. 

\subsection{Circular Buffer}
\label{sec:circular}



All secure enclave applications are partitioned into trusted \textit{enclave} code and \textit{untrusted} application code. The trusted enclave code can access encrypted memory, but cannot issue system calls; the reverse is true for untrusted application code. The boundary of this application-enclave partition is marked by \textit{ecall}s, and \textit{ocall}s. 
In order to transfer the data crossing the application-enclave boundary, a standard and straightforward approach is to invoke \textit{ecall}s and \textit{ocall}s directly, which is adopted by popular SGX container framework GrapheneSGX~\cite{tsai2017graphene}, and even enclave runtime environment Asylo~\cite{asylosocket}. The untrusted application establishes a socket and uses \textit{send} and \textit{recv} to pass messages on behalf of the enclave code. However, this approach incurs extremely high overhead. The high cost of a context switch is coupled with byte-wise copying the buffer in and out (contrasted with zero-copying). An \textit{ecall} usually takes 8,000 to 20,000 CPU cycles, and an \textit{ocall} usually takes 8,000 CPU cycles on average.

\algname enables efficient application-enclave communication by
leveraging a circular buffer data structure.
\algname initializes application and enclave by allocating two
single-publisher single-consumer circular buffers (one for each
communication direction) in untrusted plain-text memory. The memory
addresses of the allocated buffers are sent to the enclave by
{ecalls}. The circular buffer contains a number of slots, each of
them containing a pointer to the data and its size. Because the enclave
allocates and uses encrypted pages on EPC, the out-of-enclave
application cannot read the content directly given the pointer. \algname
allocates and manages several chunks of free plaintext memory and uses
pointers to the plaintext memory chunk for the circular buffer.


Using a circular buffer avoids unnecessary ecall and ocall context switches, and, unlike switchless calls \cite{openenclaveswitchless}, the writers also do not need to wait for readers to finish reading the message. This allows the writers to write multiple messages concurrently and asynchronously to the circular buffer. 

%
%


\section{PSL with \algname}
\label{sec:pslscl}

We discuss the experience and implementation effort to use SCL for PSL. 
Every PSL worker is started with a Worker Enclave in SCL, and attested by In-Enclave FaaS Leader. The code for PSL is directly executed on sandboxed Javascript engine. 
Our key distribution and management protocol provides every worker enclave with unique private keys derived from a master key by the FaaS Leader. The keys can be easily generated, verified and rotated to prevent potential key leakage.


\subsection{Sandboxing}
\label{sec:faas}
To isolate in-enclave applications from the PSL infrastructure, we use 
a sandboxed Javascript interpreter, Duktape, to dynamically interpret the Lambda program at runtime. In order for sandboxed Javascript program communicate with its other counterparts, we modify the Duktape and introduce two functions \texttt{put} and \texttt{get} to interact with SCL. We note that the program is transparent with and sandboxed from the underlying cryptographic schemes, so that it cannot observe and unintentionally leak the cryptographic secrets. 

\subsection{Attestation}
\label{sec:attestation}

\begin{figure}
    \centering
    \includegraphics[width=\linewidth]{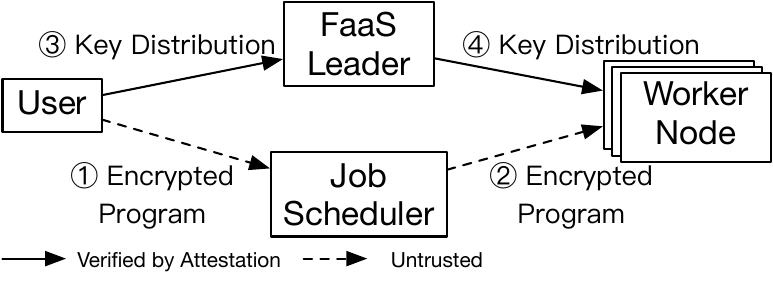}
    \caption{The launching procedure of PSL. Users dispatch encrypted programs to the job scheduler and to the worker nodes, and distribute cryptographic keys with a separate and secure channel. The channel is verified by attestation. }
    \label{fig:attestation_protocol}
\end{figure}
PSL builds its attestation protocol on top of the Asylo's attestation primitives. 
For each worker or FaaS leader that requires code running in the enclave, it starts with an Assertion Generation Enclave(AGE) as a Quoting Enclave(QE) that helps generates \textit{quotes} on behalf of the enclave. The QE is certified by the Provisioning Certification Enclave (PCE), which uses Provisioning Certification Key (PCK) that is written, and distributed by Intel to sign QE's hardware REPORT. The PCK certificate chain can be traced back to Intel SGX Root Certificate Authority(CA). After receiving an assertion request from a remote attester, the worker or FaaS leader establishes bi-directional local attestation with AGE to forward the assertion request from the remote attester and to get the assertion from the AGE. After the remote attester verifies the assertion, they establish a secure gRPC channel and the remote attester sends confidential information, such as crypographic keys, to the worker or FaaS leader. 

\subsection{Launching Process}

Each PSL worker node starts a lambda runtime in the enclave, which is registered with a third-party job scheduler. To launch a PSL workload, the user contacts the job scheduler with an encrypted program and corresponding launching configurations, such as how many lambdas are needed. The job scheduler contacts idle worker nodes within its registry and forwards the encrypted program to the potential worker nodes. To prevent malicious worker nodes, the user sends cryptographic keys via a separate channel through FaaS leader that runs in an enclave. After verifies the identity of the FaaS leader using remote attestation, the worker distributes the keys to the FaaS leader. The workers which receive the encrypted program also verify itself with remote attestaion with the FaaS leader. After the workers are authenticated, the FaaS leader forwards the cryptographic keys to the worker nodes, and the worker nodes can decrypt and run the program. When the PSL workload is finished, all the user-related confidential information, such as the content of the memtable, is cleared by a RESET command by the FaaS leader, because restarting the lambda runtime may take longer time. The FaaS leader keeps track of the idleness of the workers and only distribute keys to the idle workers. The workers after RESET need to be re-attested for the next PSL workload. 



\subsection{Key Management}
\label{sec:key_management}
In PSL, key management is needed for worker enclaves to verify each other's identity, and to satisfy the security guarantees of DataCapsules. 
Our key management design goals are: 1) Provenance: by providing a \emph{unique} key pair per worker enclave; 2) Authentication: each worker enclave needs to sign with the (derived) DataCapsule \emph{owner} identity; 3), PSL uses a hierarchical structure with a parent FaaS Leader and multiple child Lambda Enclaves. We want to design a key management scheme to efficiently manage hierarchically structured key pairs with low overhead.

To derive a each set of public/private key pairs from a master key,  we use Hierarchical Deterministic (HD) Wallet from Bitcoin Wallet\cite{nakamoto2019bitcoin}. HD Wallet is a key management scheme that allows all the child public keys to be derived from a single parent public key. We use hardened derived child keys, a scheme of HD wallet to prevent the problem of HD Wallet that the leakage of the child private key leaks the private key of the parent.  
HD Wallet enables efficient key management in PSL as follows: 1) After attestation between the client and the FaaS Leader, the client sends its owner key to the FaaS Leader. 2) The FaaS Leader generates a child public/private key pair for the current running application. 3) The FaaS Leader uses the application child key pair to generate multiple grandchild key pairs, one per worker enclave. 4) The FaaS Leader attests and sends every enclave its grandchild key pair. 5) FaaS Leader multicasts the application public key to all enclaves. 6) Each worker enclave derives the other worker enclaves' public keys using the application public key.

With this key management scheme, both provenance and authentication are achieved. In particular, 1) every worker enclave has its own signing key (\emph{i.e.} provenance), and 2) every worker can sign messages on behalf of the owner identity using a derived grandchild key pair (\emph{i.e.} authentication). This scheme minimizes key exchanges among the client, the FaaS Leader, and worker enclaves. For $n$ worker enclaves, the initial key exchange overhead reduces from possibly $O(n^2)$ for a naive key management scheme to $O(n)$. 

\paragraph{Key Leakage and Rotation}
\label{sec:key_leakage_and_rotation}
We enable efficient key rotation scheme with SCL that can derive and distribute a new set of key pairs for the workers from the new hardened key pair. This prevents the cryptographic key leakage over time. This is done by (1) client deriving a new child hardened key pair and multicasting the public key to all enclave workers; (2) the FaaS Leader then derives a new set of key pairs for the workers from the new key pair. 
To handle lost multicasted messages or enclave worker failure, we can rely on SCL's consistency coordinator and include the current parent public key in the SYNC reports. This ensures that any enclave worker can verify that they are using the correct signing keys in a given epoch by validating the keys against the consistency coordinator's SYNC reports. The frequency in which key rotation occurs depends on the user's threat model. Users may choose to rotate keys per function invocation. This ensures any new function invocations may not affect previous function invocations. 
\section{Implementation}

Our codebase contains 32,454 LoC in C++ excluding comments and 43,011 LoC code base in total counted by cloc\cite{cloc}. The core SCL KVS code consisted of roughly 4,000 lines of code in C++, excluding the attestation, distributive application implementations, and experiment scripts. We implement the KVS directly on top of Asylo instead of on a containerized enclave environment. This yields a much smaller TCB than related works such as Speicher \cite{bailleu2019speicher}.  

Asylo is a hardware-agnostic framework for TEEs, supporting Intel SGX(v1 and v2) and ARM TrustZone. It also provided a POSIX compliant library that made it easier to port existing applications into enclaves. We use ZeroMQ to implement network multicast and communications between Worker Enclaves. We use gRPC to create a secure FaaS Leader Enclave, which can generate HD Wallet keypairs and startup enclave workers.  We use DukTape, an embedded JavaScript engine in C++, to sandbox enclave applications, now that enclaves can directly execute JavaScript code. 

CapsuleDB is implemented in C++ and is ~2200 LoC.  It also uses several features of Asylo and the structures created in the PSL implementation.  We use a similar memtable implementation, but leverage mutexes on each entry instead of a spinlock.  Due to the implementation timeline, the current version of CapsuleDB writes data to disc rather than to a network attached DataCapsule using the Boost serialization library.  The DataCapsule replication service contains about 1,000 LoC in C++ excluding comments. We use RocksDB as embedded persistent storage for each DataCapsule replica, ZeroMQ to implement network communication between DataCapsule replicas, and OpenSSL for signature and verification.

\section{Evaluation}
\algname leverages DataCaspules as the data representation to support inter-enclave communication. To quantify the benefits and limitations, we ask: 
(1) How does \algname perform as a KVS(\S\ref{sec:e2eb})? 
(2) How do circular buffer (\S \ref{sec:circularb}), and replication (\S \ref{sec:dcr_e2e}) affect the overhead?  (3) How long does it take to securely launch a PSL task? (\S \ref{sec:lambda_launch}) (4) How much does \algname pay to run in-enclave distributed applications(\S \ref{sec:mplambda})?

\subsection{Experiment Setup}
We evaluate PSL on fifteen Intel NUCs 7PJYH, equiped with Intel(R) Pentium(R) Silver J5005 CPU @ 1.50GHz with 4 physical cores (4 logical threads). The processor has 96K L1 data cache, a 4MiB L2 cache, and 16GB memory. The machine uses Ubuntu 18.04.5 LTS 64bit with Linux 5.4.0-1048-azure. We run Asylo version 0.6.2. We report the average of experiments that are conducted 10 times. For each NUC, it runs two PSL threads by default. 

\subsection{End-To-End Benchmark of \algname}
\label{sec:e2eb}
\paragraph{Benchmark Design:} An end-to-end evaluation of \algname starts the worker sends the first acknowledgement to the user, and ends when the client receives its last request's response from the workers. We evaluate the performance using a workload generated by YCSB workload generators. Due to the difference between \texttt{get} and \texttt{put} protocols, we focus on the read-only and write-only workloads. All workloads comply zipfian distribution, where keys are accessed at non-uniform frequency. For each \texttt{get}, we evaluate the performance of getting from the local memtable of the lambda(\texttt{get(cached)}), and of getting the data from CapsuleDB(\texttt{get(uncached)}). Each \texttt{get} request is synchronous that the next request is sent only if it gets the value of the previous \texttt{get} request. 


\begin{figure}
    \centering
    \includegraphics[width=0.47\linewidth]{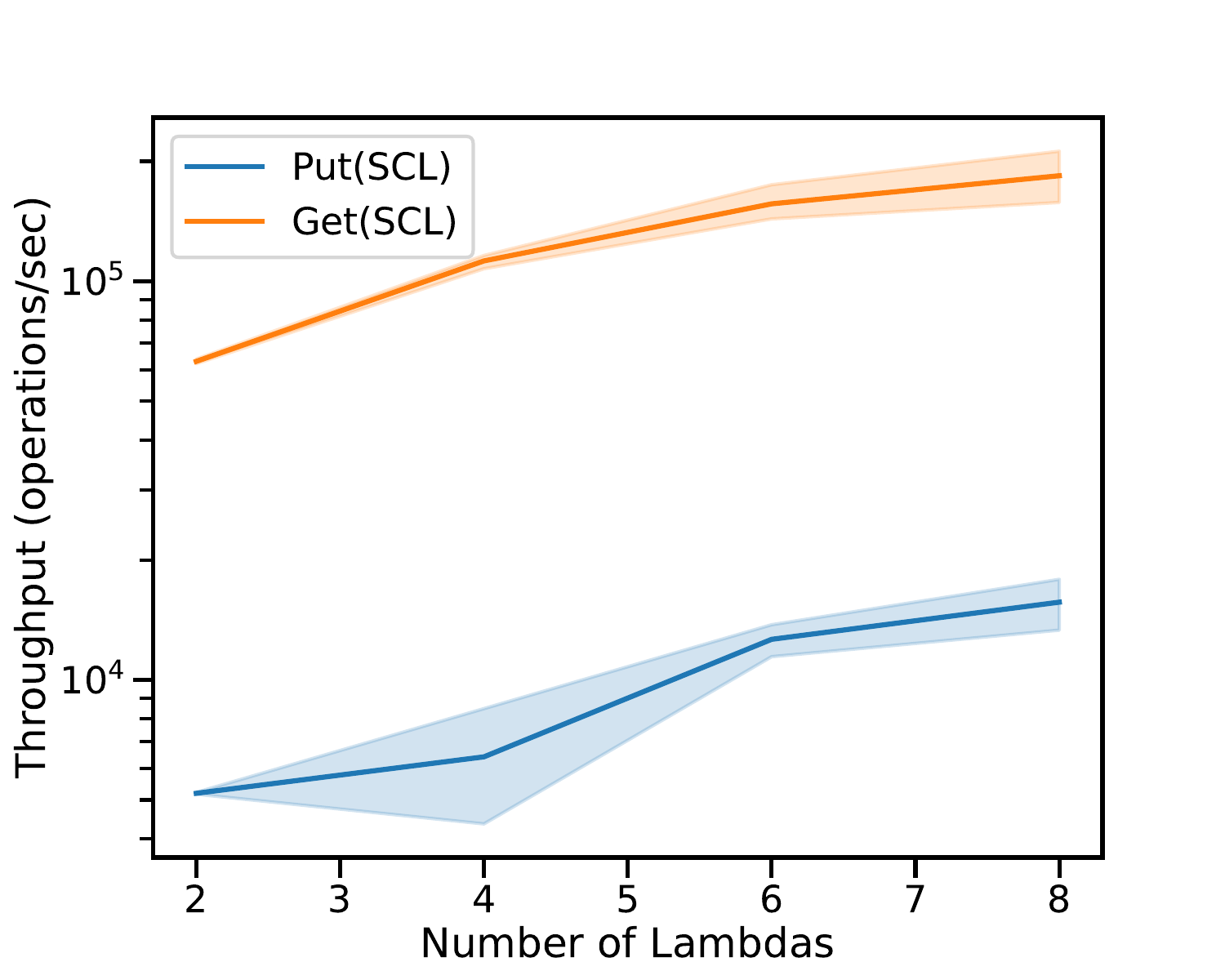}
    \includegraphics[width=0.47\linewidth]{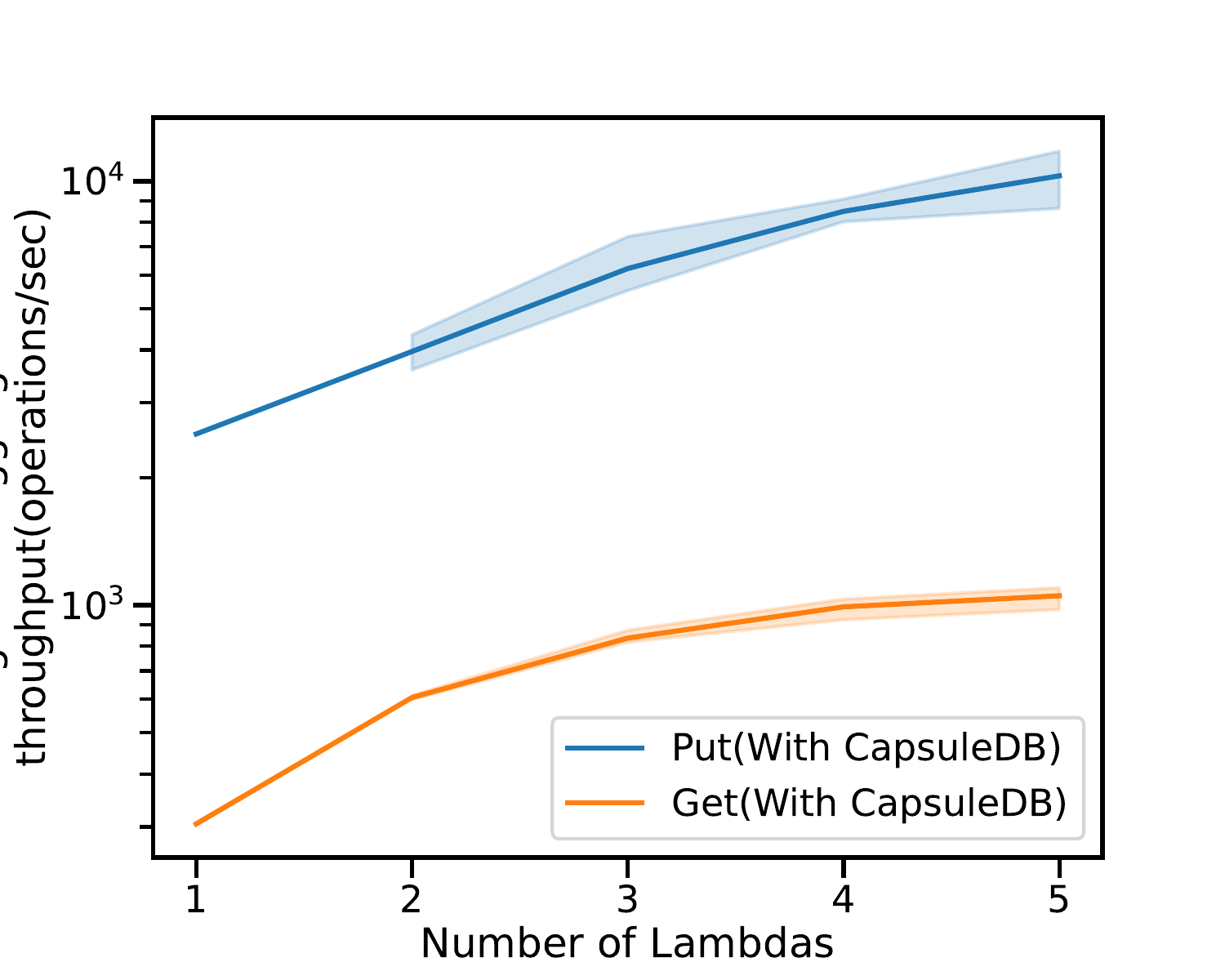}
    \caption{The aggregated throughput of PSL on put-only and get-only workload of YCSB benchmark. The \texttt{get}s are differentiated by whether it is cached in the memtable(left) or the lambda needs to query CapsuleDB(right). }
    \label{fig:e2eb}
\end{figure}

\paragraph{Overall Performance:} 
Figure \ref{fig:e2eb} shows the throughput of the end-to-end YCSB benchmark. The aggregated throughput of \texttt{put}. The \texttt{get(CapsuleDB)} throughput is flattened as we increate the number of the lambdas, because we run one single CapsuleDB instance that handle all the queries, which is bottlenecked as the number of lambdas that issue \texttt{get(CapsuleDB)} increases.

\subsection{Replication-enabled End-To-End Benchmark}
\label{sec:dcr_e2e}
\paragraph{Benchmark Design:} Replication-enabled end-to-end evaluation measures the performance of the SCL layer with durability. In particular, it includes the overhead of workers sending each write to the DataCapsule replicas, a quorum of DataCapsule replicas receive data and persist it on disk, and then acking the worker. We evaluate the performance using a workload generated by YCSB workload generators. Since replication involves only write operations, we evaluate a write-only workload. The workload involves a zipfian distribution, with keys accessed at a non-uniform frequency. 

\begin{figure}
\vspace*{-0.5 cm}
    \centering
    \includegraphics[width=0.8\linewidth]{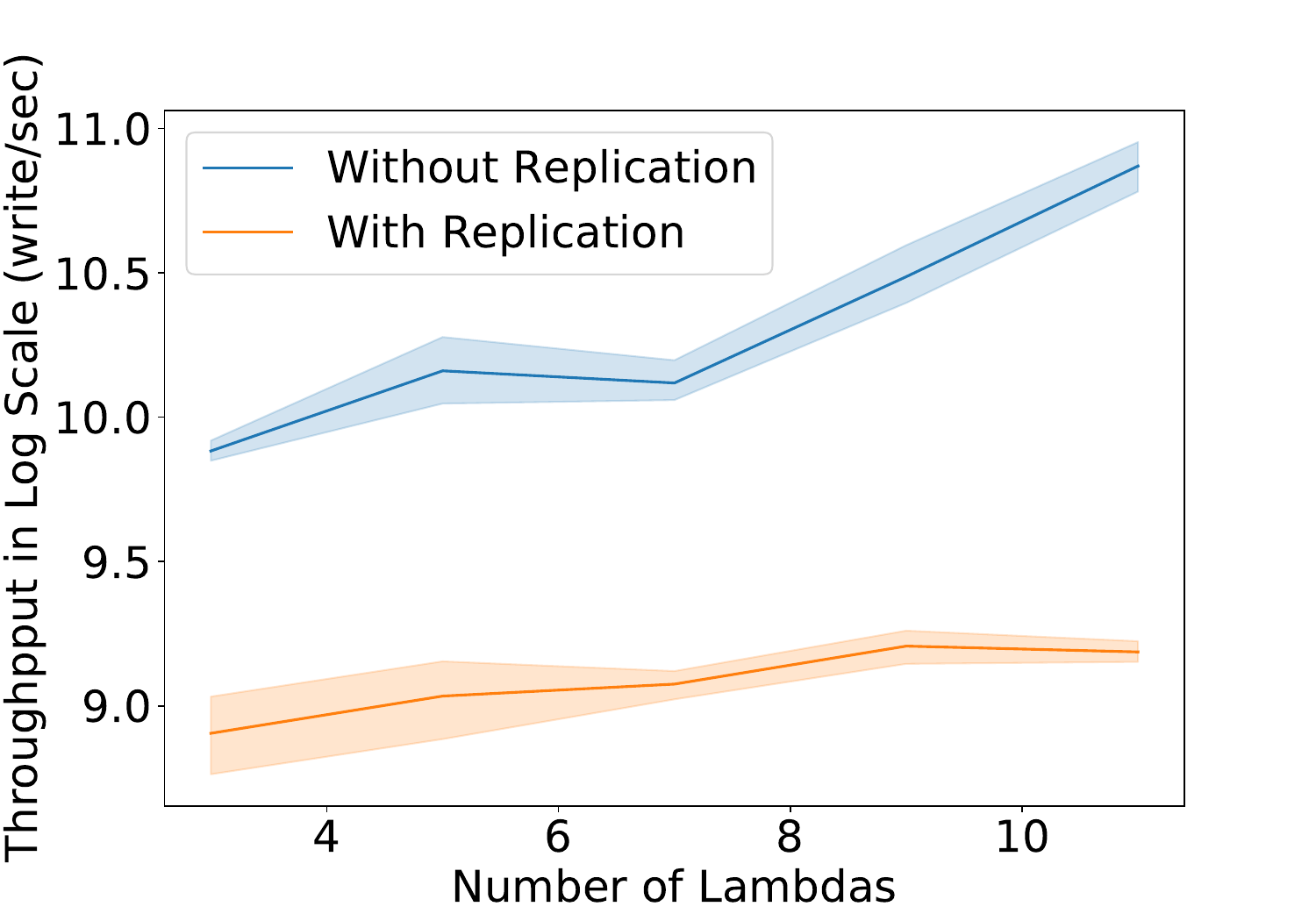}
    \caption{A line graph that shows end-to-end write-only benchmarks for SCL with vs. without replication. Throughput numbers are in log scale.}
    \label{fig:dcr_e2e}
\end{figure}

\paragraph{Overall Performance:} Figure \ref{fig:dcr_e2e}
illustrates the performance of the DataCapsule backend.  It shows that SCL with replication has reached a bottleneck after 9 workers while SCL without durability continues to scale. The performance drop and bottleneck are due to several reasons: 1) disk operations are inherently slow; 2) the burden on replication system's leader is high for collecting acks from DataCapsule replicas and sending the aggregated ack back to worker. We aim to improve SCL with replication by mitigating the workload on the replication leader.

\subsection{Circular Buffer Microbenchmark}
\label{sec:circularb}
\paragraph{Benchmark Design:} The circular buffer provides efficient application-enclave communication. We compare the performance of the circular buffer with the SGX SDK baseline and the state-of-the-art HotCall. We evaluate them based on the number of clock cycles required for communications in both the application to enclave direction and vice versa.

\paragraph{Overall Performance:} As shown in Table \ref{table:microbenchmark}, baseline SGX SDK incurs a significant overhead of over 20,000 clock cycles from application to enclave, and over 8,600 clock cycles from enclave to application. For both directions, HotCall is able to reduce the overhead to under a thousand clock cycles. Our circular buffer reduces overheads even further. Our solution only requires 461.1 and 525.54 clock cycles from application to enclave and vice versa. Compared to state-of-the-art HotCall, our solution provides 103\% and 44\% improvements, respectively.

\begin{table}
  \begin{tabular}{| p{0.15\textwidth} | p{0.13\textwidth} | p{0.13\textwidth} |}
    \hline
     \textbf{\# of clock cycles}  & \textbf{App to Enclave} & \textbf{Enclave to App} \\
    \hline
    \textbf{SGX SDK} & 20515.02 & 8608.57 \\
    \hline
    \textbf{HotCalls} & 936.89 & 757.96 \\
    \hline
    \textbf{Circular Buffer} & 461.10 & 525.54  \\
    \hline
\end{tabular}
\caption{\label{table:microbenchmark} \textbf{Circular Buffer Microbenchmark} We evaluate the number of clock cycles required for communications between the enclave and application in both directions.}
\end{table}

\begin{figure}
    \centering
    \includegraphics[width=\linewidth]{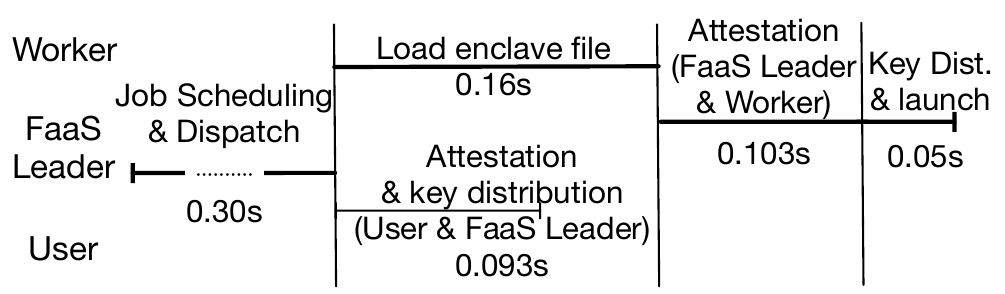}
    \caption{Latency breakdown of the Paranoid Stateful Lambda launching process. The bold line represents the critical path of the lambda launching process. The total launching time to run code in authenticated worker is less than 0.61 second. }
    \label{fig:attestation_latency}
\end{figure}

\subsection{Lambda Launch Time} 
\label{sec:lambda_launch}
\paragraph{Benchmark Design:} We evaluate the launching process of PSL by running Workers and FaaS leader in SGXv2 hardware mode, which the worker lambda, FaaS leader and user on different physical Intel NUCs machines. For each NUC, it runs Asylo AGE in hardware mode with PCE signed by Intel that helps enclave generates attestation assertions. We assume the machines already have the pulled the prebuilt lambda runtime binaries and execute the runtime. The cold-start bootstrapping process lasts 42 seconds on average in our experiment setting. 

\paragraph{Lambda Launch Breakdown:}
Figure \ref{fig:attestation_latency} show the latency breakdown of the launching process. It takes 0.30s for the user to reach out to the scheduler, and for the scheduler to find and forward the encrypted task to the potential workers. Then the workers load associated runtime and data to the enclave, which takes 0.16s. We parallelize the worker loading time with the attestation. that the user remotely attests the FaaS Leader to verify that the FaaS leader is running authenticated code in SGX enclave. After the worker's enclave file is loaded, it takes 0.103s on average for the FaaS leader to remotely attest the worker enclave. We note that this attestation latency is mostly constituted by the network delay of grpc request and the local attestation assertion generation time of the worker's AGE, so it does not incur scalability issue with the FaaS leader when multiple workers are launched at the same time.

\subsection{Case Study: Fog Robotics Motion Planner}
\label{sec:mplambda}

We experiment with a sampling-based motion planner that is parallelized to run on multiple concurrent serverless processes, \texttt{MPLambda}~\cite{ICRAJeffIcknowski2020}, and modifying it to use SCL. 
Most of the porting effort done was to integrate \texttt{MPLambda}'s build system into Asylo. The modification is about 100 LoC. Many system calls that \texttt{MPLambda} uses are proxied by Asylo. 

\begin{table}[t]
\centering
  \begin{tabular}{| c | c | c | c | c |}
    \hline
    \textbf{No.}  & \textbf{TTFS} & \textbf{TTFS} & \textbf{Cost/Time} & \textbf{Cost/Time}  \\
    \textbf{Planners}  & \textbf{Baseline} & \textbf{SCL} & \textbf{Baseline} & \textbf{SCL}  \\
    \hline
    \textbf{1} & 61.9 sec & TIMEOUT & 0.30 & N/A \\
    \hline
    \textbf{2} & 115.5 sec & TIMEOUT	& 0.16 & N/A \\
    \hline
    \textbf{4} & 89.4 sec & 481.5 sec  & 0.09 & 0.77 \\
     \hline
    \textbf{8} & 34.7 sec & 207.9 sec & 0.08 & 0.13\\
    \hline
\end{tabular}
\caption{Motion Planning Benchmarks} 
\vspace*{-0.1in}
\end{table}

Using MPLambda with SCL, we compute a motion plan running a fetch scenario in which a Fetch mobile manipulator robot~\cite{Fetch} declutters a desk. We measure the median wall-clock time to find the first solution by the planners. We also measure the median average path cost per time of the lowest cost path the planners return. This captures how efficiently the planners can compute the best path. Because the planner uses random sampling, we run the same computation multiple times with different seeds. As with previous experiments, we run this test on Intel NUCs 7PJYH, equipped with Intel(R) Pentium(R) Silver J5005 CPU @ 1.50GHz with 4 physical cores (4 logical threads). We set a timeout of 600 seconds for the planners to compute a path. 

We run up to 8 planners, running on separate Intel NUCs using SCL and comparing this to running MPLamda without SCL. We observe an increase in performance as we scale out the number of planners. Each planner runs computationally heavy workloads and PSL introduces several threads (i.e. crypto actors, zmq clients, OCALL/ECALL handlers) that take away CPU time from the planner thread. Furthermore, MPLambda planners use the Rapidly-exploring random tree (RRT*) \cite{karaman2011samplingbased} algorithm, to search for paths by randomly generating samples from a search space, checking whether the sample is feasible to explore, and adding the sample to a constructed tree data structure. The tree data structure may grow large and take up a significant amount of memory. Memory in SGX is a limited resource and increased memory pressure leads to more misses in the EPC and requiring paging in and out of enclaves frequently. There is work on limiting the memory usage of RRT* by bounding the memory for the tree data structure, which we can adopt in future work. \cite{10.1109/ICMA.2013.6617944}.


\section{Related Work}
\paragraph{Current Frameworks for FaaS:}
Existing cloud-based FaaS implementations, such as AWS Lambda \cite{lambda} or OpenFaaS \cite{OpenFaaS}, underutilize computing resources on the edge of the network. Attempts to deploy such frameworks to the edge, such as Akamai \cite{akamai}, do not deliver the security guarantee required by the Edge Computing. S-FaaS \cite{alder2019s}, Clemmys \cite{trach2019clemmys} uses TEE  and cryptographic attestation to protect the confidentiality of the execution.  For all the aforementioned FaaS frameworks, they do not support stateful FaaS execution \cite{sreekanti2020cloudburst}. 



\paragraph{Secure Execution with TEE:} 
PSL is motivated by the vision that the distributive worker can run securely in a TEE on a single host, making the security and efficiency of communication among multiple enclaves a logical research problem. This vision is supported by a variety of available container services and platforms, for example, TEE-enabled container services such as GrapheneSGX \cite{tsai2017graphene},  Scone \cite{arnautov2016scone}, and Occlum \cite{shen2020occlum} and hardware TEE platforms \cite{lee2019keystone} , Elasticlave \cite{yu2020elasticlave} and Penglai \cite{fengscalable}. Snort \cite{kuvaiskii2018snort} is an in-enclave intrusion detection framework that also uses a circular buffer for communication. We note our approach differs from Snort in that they use circular buffers to convert hugepages in DPDK, while our circular buffer design is to eliminate the context switch in ecalls/ocalls.


\paragraph{KVS based on TEE:} 
Existing TEE-based KVS designs mainly focus on single-TEE persistence and performance optimizations. ShieldStore \cite{kim2019shieldstore} solves the 128MB limitation of SGXv1 by conducting most processing outside the enclave. Each key-value pair is encrypted and protected with a signature when it leaves the enclave, and the main data structures of the KVS are also stored outside the enclave. The in-enclave KVS server handles queries from an out-of-enclave client by fetching encrypted key-value pairs from untrusted memory.
Speicher \cite{bailleu2019speicher} and DiskShield \cite{ahn2020diskshield} implement secure storage inside a secure enclave, so that the TEE can exchange data securely to the underlying storage of the host. Both \algname and Speicher \cite{bailleu2019speicher} use a LSM-based structure for durablity, but \algname takes a step further to integrate the stored data blocks as part of the DataCapsule hash chain, and to enable efficient inter-enclave communication. \algname also has a much smaller TCB required than Speicher. 
EnclaveCache \cite{10.1145/3361525.3361533} and Omega \cite{correia2020omega} supports shared, in-memory KVS cache but does not support communication of enclaves from different hosts. 

\section{Conclusion}

We introduced Paranoid Stateful Lambdas, a federated FaaS framework for secure and stateful execution in both cloud and edge computing environments. We focus on the security and communication aspects of PSL by exploiting the properties and extensions of DataCaspules, a cryptographically-hardened blockchain. We propose an abstraction, the Secure Concurrency Layer, that provides security and eventual consistency to the enclaves, as well as discuss its durability and fault tolerance semantics. 
On our end-to-end benchmark, SCL has up to 81x higher throughput and 2.08x lower latency than the unoptimized baseline. 
Our system throughput scales linearly with the number of the lambdas, and our lambda task can be dispatched to authenticated workers within 0.61 second. 

\section* {Acknowledgment}
We thank Anoop Jaishankar for great discussion on Asylo attestation. 
This material is based upon work supported by NSF/VMware Partnership on Edge
Computing Data Infrastructure (ECDI), NSF award 1838833.
Any opinions, findings, and conclusions or recommendations
expressed in this material are those of the authors and do not
necessarily reflect the views of the sponsors

\bibliographystyle{plain}
\bibliography{main}

\end{document}